\documentclass[12pt]{article}
\usepackage{epsfig,a4}

\def\ga{\mathrel{\raise.3ex\hbox{$>$\kern-.75em\lower1ex\hbox{$\sim$}}}}
\def\la{\mathrel{\raise.3ex\hbox{$<$\kern-.75em\lower1ex\hbox{$\sim$}}}}

\newcommand{\Tcsq}{\ensuremath{\mathrm{TeV}/c^2}}
\newcommand{\Gcsq}{\ensuremath{\mathrm{GeV}/c^2}}

\newcommand{\rts}{\ensuremath{\sqrt{s}}}
\newcommand{\lumi} {\mathcal{L}}
\newcommand{\Lumi} {\int\lumi\mathrm{d}t}
\newcommand{\pbinv}{\ensuremath{\mathrm{pb}^{-1}}}

\newcommand{\epem}{\ensuremath{\mathrm{e^+e^-}}}

\newcommand{\gaga}{\ensuremath{\gamma\gamma}}
\newcommand{\gagahh}{\ensuremath{\gamma\gamma\rightarrow{\mathrm{hadrons}}}}
\newcommand{\ggll}{\ensuremath{\gamma\gamma\rightarrow{\mathrm{leptons}}}}
\newcommand{\gghh}{\gagahh}

\newcommand{\wen}{\ensuremath{\mathrm{We\nu}}}
\newcommand{\znn}{\ensuremath{\mathrm{Z\nu\bar{\nu}}}}
\newcommand{\ww}{\ensuremath{\mathrm{W^+W^-}}}
\newcommand{\wwll}{\ensuremath{\ww\to \ell^+\nu \ell^-\bar{\nu}}}

\newcommand{\tanb}{\ensuremath{\tan\beta}}

\newcommand{\mzero}{\ensuremath{m_{0}}}

\newcommand{\At}{\ensuremath{A_{\mathrm t}}}
\newcommand{\Azero}{\ensuremath{A_0}}
\newcommand{\Pcha}{\ensuremath{\chi^{\pm}}}
\newcommand{\Pchap}{\ensuremath{\chi^{\pm}_2}}

\newcommand{\PChiz}[1]{\ensuremath{\chi^0_{#1}}}
\newcommand{\PSnu}{\ensuremath{\widetilde{\nu}}}

\newcommand{\PSl}{\ensuremath{\widetilde{\ell}}}

\newcommand{\PStaur}{\ensuremath{\mathrm{\tilde{\tau}}_\mathrm{R}}}

\newcommand{\PSe}{\ensuremath{\widetilde{\mathrm{e}}}}
\newcommand{\PSmu}{\ensuremath{\widetilde{\mu}}}
\newcommand{\PStau}{\ensuremath{\widetilde{\tau}}}

\newcommand{\PSb}{\ensuremath{\mathrm{\widetilde{b}}}}

\newcommand{\PSbl}{\ensuremath{\mathrm{\tilde{b}}_\mathrm{L}}}
\newcommand{\PSt}{\ensuremath{\mathrm{\widetilde{t}}}}
\newcommand{\PSq}{\ensuremath{\mathrm{\widetilde{q}}}}
\newcommand{\PSgl}{\ensuremath{\mathrm{\widetilde{g}}}}

\newcommand{\Pchi}{\ensuremath{\chi}}

\newcommand{\Mchi}{\ensuremath{m_{\PChiz{1}}}}

\newcommand{\mA}{\ensuremath{m_{\mathrm A}}}
\newcommand{\mh}{\ensuremath{m_{\mathrm h}}}

\newcommand{\mt}{\ensuremath{m_{\mathrm t}}}

\newcommand{\dm}{\ensuremath{\Delta M}}

\newcommand{\neui}{\ensuremath{\chi^0_{i}}}
\newcommand{\neuj}{\ensuremath{\chi^0_{j}}}

\newcommand{\neuu}{\ensuremath{\chi^0_{1}}}

\newcommand{\thetat}{\ensuremath{\theta_{\PSt}}}
\newcommand{\thetab}{\ensuremath{\theta_{\PSb}}}

\font\ninerm=cmr9

\begin{document}
\thispagestyle{empty}

\begin{picture}(160,1)
\put(-2,15){\rm {ORGANISATION EUROP\'EENNE POUR LA RECHERCHE NUCL\'EAIRE (CERN)}} 
\put(28,10){\rm Laboratoire Europ\'een pour la Physique des Particules}
\put(115,-10) {\parbox[t]{45mm}{\tt CERN-EP/2000-139}}
\put(115,-15){\parbox[t]{45mm}{\tt 13\ November\ 2000}}
\end{picture}

\vskip 3cm

\begin{center}
{\LARGE
Search for Supersymmetric Particles \\
in $\epem$ Collisions at $\sqrt{s}$ up to 202\,GeV \\
and \\
Mass Limit for the Lightest Neutralino\\
}
\vskip 1cm

{\large The ALEPH Collaboration$^*)$}
\end{center}

\vskip 3.cm

\begin{abstract}
Searches for pair production of squarks, sleptons, charginos and 
neutralinos are performed with the data collected by the ALEPH detector 
at LEP at centre-of-mass energies from 188.6 to 201.6\,GeV.
No evidence for any such signals is observed in a total 
integrated luminosity of about 410\,\pbinv. 
The negative results of the searches are translated into exclusion 
domains in the space of the relevant MSSM parameters, 
which improve significantly on the constraints set previously. 
Under the assumptions of gaugino and sfermion mass unification, 
these results allow a 95\%~C.L. 
lower limit of 37\,\Gcsq\ to be set on the mass of the 
lightest neutralino for any \tanb\ and sfermion mass.
Additional constraints in the MSSM parameter space are derived 
from the negative results of ALEPH searches for Higgs bosons. 
The results are also interpreted in the framework of minimal supergravity.
\end{abstract}

\vfill
\centerline{\it To be submitted to Physics Letters B}
\vskip .5cm
\noindent
--------------------------------------------\hfil\break
{\ninerm $^*)$ See next pages for the list of authors}

\eject



\pagestyle{empty}
\newpage
\small
%
%
\newlength{\saveparskip}
\newlength{\savetextheight}
\newlength{\savetopmargin}
\newlength{\savetextwidth}
\newlength{\saveoddsidemargin}
\newlength{\savetopsep}
\setlength{\saveparskip}{\parskip}
\setlength{\savetextheight}{\textheight}
\setlength{\savetopmargin}{\topmargin}
\setlength{\savetextwidth}{\textwidth}
\setlength{\saveoddsidemargin}{\oddsidemargin}
\setlength{\savetopsep}{\topsep}
%
%
\setlength{\parskip}{0.0cm}
\setlength{\textheight}{25.0cm}
\setlength{\topmargin}{-1.5cm}
\setlength{\textwidth}{16 cm}
\setlength{\oddsidemargin}{-0.0cm}
\setlength{\topsep}{1mm}
\pretolerance=10000
\centerline{\large\bf The ALEPH Collaboration}
\footnotesize
\vspace{0.5cm}
{\raggedbottom
\begin{sloppypar}
\samepage\noindent
R.~Barate,
I.~De~Bonis,
D.~Decamp,
P.~Ghez,
C.~Goy,
S.~Jezequel,
J.-P.~Lees,
F.~Martin,
E.~Merle,
\mbox{M.-N.~Minard},
B.~Pietrzyk
\nopagebreak
\begin{center}
\parbox{15.5cm}{\sl\samepage
Laboratoire de Physique des Particules (LAPP), IN$^{2}$P$^{3}$-CNRS,
F-74019 Annecy-le-Vieux Cedex, France}
\end{center}\end{sloppypar}
\vspace{2mm}
\begin{sloppypar}
\noindent
S.~Bravo,
M.P.~Casado,
M.~Chmeissani,
J.M.~Crespo,
E.~Fernandez,
M.~Fernandez-Bosman,
Ll.~Garrido,$^{15}$
E.~Graug\'{e}s,
J.~Lopez,
M.~Martinez,
G.~Merino,
R.~Miquel,
Ll.M.~Mir,
A.~Pacheco,
D.~Paneque,
H.~Ruiz
\nopagebreak
\begin{center}
\parbox{15.5cm}{\sl\samepage
Institut de F\'{i}sica d'Altes Energies, Universitat Aut\`{o}noma
de Barcelona, E-08193 Bellaterra (Barcelona), Spain$^{7}$}
\end{center}\end{sloppypar}
\vspace{2mm}
\begin{sloppypar}
\noindent
A.~Colaleo,
D.~Creanza,
N.~De~Filippis,
M.~de~Palma,
G.~Iaselli,
G.~Maggi,
M.~Maggi,$^{1}$
S.~Nuzzo,
A.~Ranieri,
G.~Raso,$^{24}$
F.~Ruggieri,
G.~Selvaggi,
L.~Silvestris,
P.~Tempesta,
A.~Tricomi,$^{3}$
G.~Zito
\nopagebreak
\begin{center}
\parbox{15.5cm}{\sl\samepage
Dipartimento di Fisica, INFN Sezione di Bari, I-70126 Bari, Italy}
\end{center}\end{sloppypar}
\vspace{2mm}
\begin{sloppypar}
\noindent
X.~Huang,
J.~Lin,
Q. Ouyang,
T.~Wang,
Y.~Xie,
R.~Xu,
S.~Xue,
J.~Zhang,
L.~Zhang,
W.~Zhao
\nopagebreak
\begin{center}
\parbox{15.5cm}{\sl\samepage
Institute of High Energy Physics, Academia Sinica, Beijing, The People's
Republic of China$^{8}$}
\end{center}\end{sloppypar}
\vspace{2mm}
\begin{sloppypar}
\noindent
D.~Abbaneo,
P.~Azzurri,
T.~Barklow,$^{30}$
G.~Boix,$^{6}$
O.~Buchm\"uller,
M.~Cattaneo,
F.~Cerutti,
B.~Clerbaux,
G.~Dissertori,
H.~Drevermann,
R.W.~Forty,
M.~Frank,
F.~Gianotti,
T.C.~Greening,
J.B.~Hansen,
J.~Harvey,
D.E.~Hutchcroft,
P.~Janot,
B.~Jost,
M.~Kado,
V.~Lemaitre,
P.~Maley,
P.~Mato,
A.~Minten,
A.~Moutoussi,
F.~Ranjard,
L.~Rolandi,
D.~Schlatter,
M.~Schmitt,$^{20}$
O.~Schneider,$^{2}$
P.~Spagnolo,
W.~Tejessy,
F.~Teubert,
E.~Tournefier,$^{26}$
A.~Valassi,
J.J.~Ward,
A.E.~Wright
\nopagebreak
\begin{center}
\parbox{15.5cm}{\sl\samepage
European Laboratory for Particle Physics (CERN), CH-1211 Geneva 23,
Switzerland}
\end{center}\end{sloppypar}
\vspace{2mm}
\begin{sloppypar}
\noindent
Z.~Ajaltouni,
F.~Badaud,
S.~Dessagne,
A.~Falvard,
D.~Fayolle,
P.~Gay,
P.~Henrard,
J.~Jousset,
B.~Michel,
S.~Monteil,
\mbox{J-C.~Montret},
D.~Pallin,
J.M.~Pascolo,
P.~Perret,
F.~Podlyski
\nopagebreak
\begin{center}
\parbox{15.5cm}{\sl\samepage
Laboratoire de Physique Corpusculaire, Universit\'e Blaise Pascal,
IN$^{2}$P$^{3}$-CNRS, Clermont-Ferrand, F-63177 Aubi\`{e}re, France}
\end{center}\end{sloppypar}
\vspace{2mm}
\begin{sloppypar}
\noindent
J.D.~Hansen,
J.R.~Hansen,
P.H.~Hansen,
B.S.~Nilsson,
A.~W\"a\"an\"anen
\nopagebreak
\begin{center}
\parbox{15.5cm}{\sl\samepage
Niels Bohr Institute, 2100 Copenhagen, DK-Denmark$^{9}$}
\end{center}\end{sloppypar}
\vspace{2mm}
\begin{sloppypar}
\noindent
G.~Daskalakis,
A.~Kyriakis,
C.~Markou,
E.~Simopoulou,
A.~Vayaki
\nopagebreak
\begin{center}
\parbox{15.5cm}{\sl\samepage
Nuclear Research Center Demokritos (NRCD), GR-15310 Attiki, Greece}
\end{center}\end{sloppypar}
\vspace{2mm}
\begin{sloppypar}
\noindent
A.~Blondel,$^{12}$
\mbox{J.-C.~Brient},
F.~Machefert,
A.~Roug\'{e},
M.~Swynghedauw,
R.~Tanaka
\linebreak
H.~Videau
\nopagebreak
\begin{center}
\parbox{15.5cm}{\sl\samepage
Laboratoire de Physique Nucl\'eaire et des Hautes Energies, Ecole
Polytechnique, IN$^{2}$P$^{3}$-CNRS, \mbox{F-91128} Palaiseau Cedex, France}
\end{center}\end{sloppypar}
\vspace{2mm}
\begin{sloppypar}
\noindent
E.~Focardi,
G.~Parrini,
K.~Zachariadou
\nopagebreak
\begin{center}
\parbox{15.5cm}{\sl\samepage
Dipartimento di Fisica, Universit\`a di Firenze, INFN Sezione di Firenze,
I-50125 Firenze, Italy}
\end{center}\end{sloppypar}
\vspace{2mm}
\begin{sloppypar}
\noindent
A.~Antonelli,
M.~Antonelli,
G.~Bencivenni,
G.~Bologna,$^{4}$
F.~Bossi,
P.~Campana,
G.~Capon,
V.~Chiarella,
P.~Laurelli,
G.~Mannocchi,$^{5}$
F.~Murtas,
G.P.~Murtas,
L.~Passalacqua,
M.~Pepe-Altarelli$^{25}$
\nopagebreak
\begin{center}
\parbox{15.5cm}{\sl\samepage
Laboratori Nazionali dell'INFN (LNF-INFN), I-00044 Frascati, Italy}
\end{center}\end{sloppypar}
\vspace{2mm}
\begin{sloppypar}
\noindent
M.~Chalmers,
A.W.~Halley,
J.~Kennedy,
J.G.~Lynch,
P.~Negus,
V.~O'Shea,
B.~Raeven,
D.~Smith,
P.~Teixeira-Dias,
A.S.~Thompson
\nopagebreak
\begin{center}
\parbox{15.5cm}{\sl\samepage
Department of Physics and Astronomy, University of Glasgow, Glasgow G12
8QQ,United Kingdom$^{10}$}
\end{center}\end{sloppypar}
\begin{sloppypar}
\noindent
R.~Cavanaugh,
S.~Dhamotharan,
C.~Geweniger,
P.~Hanke,
V.~Hepp,
E.E.~Kluge,
G.~Leibenguth,
A.~Putzer,
K.~Tittel,
S.~Werner,$^{19}$
M.~Wunsch$^{19}$
\nopagebreak
\begin{center}
\parbox{15.5cm}{\sl\samepage
Kirchhoff-Institut f\"ur Physik, Universit\"at Heidelberg, D-69120
Heidelberg, Germany$^{16}$}
\end{center}\end{sloppypar}
\vspace{2mm}
\begin{sloppypar}
\noindent
R.~Beuselinck,
D.M.~Binnie,
W.~Cameron,
G.~Davies,
P.J.~Dornan,
M.~Girone,$^{1}$
N.~Marinelli,
J.~Nowell,
H.~Przysiezniak,
J.K.~Sedgbeer,
J.C.~Thompson,$^{14}$
E.~Thomson,$^{23}$
R.~White
\nopagebreak
\begin{center}
\parbox{15.5cm}{\sl\samepage
Department of Physics, Imperial College, London SW7 2BZ,
United Kingdom$^{10}$}
\end{center}\end{sloppypar}
\vspace{2mm}
\begin{sloppypar}
\noindent
V.M.~Ghete,
P.~Girtler,
E.~Kneringer,
D.~Kuhn,
G.~Rudolph
\nopagebreak
\begin{center}
\parbox{15.5cm}{\sl\samepage
Institut f\"ur Experimentalphysik, Universit\"at Innsbruck, A-6020
Innsbruck, Austria$^{18}$}
\end{center}\end{sloppypar}
\vspace{2mm}
\begin{sloppypar}
\noindent
E.~Bouhova-Thacker,
C.K.~Bowdery,
D.P.~Clarke,
G.~Ellis,
A.J.~Finch,
F.~Foster,
G.~Hughes,
R.W.L.~Jones,$^{1}$
M.R.~Pearson,
N.A.~Robertson,
M.~Smizanska
\nopagebreak
\begin{center}
\parbox{15.5cm}{\sl\samepage
Department of Physics, University of Lancaster, Lancaster LA1 4YB,
United Kingdom$^{10}$}
\end{center}\end{sloppypar}
\vspace{2mm}
\begin{sloppypar}
\noindent
I.~Giehl,
F.~H\"olldorfer,
K.~Jakobs,
K.~Kleinknecht,
M.~Kr\"ocker,
A.-S.~M\"uller,
H.-A.~N\"urnberger,
G.~Quast,$^{1}$
B.~Renk,
E.~Rohne,
H.-G.~Sander,
S.~Schmeling,
H.~Wachsmuth,
C.~Zeitnitz,
T.~Ziegler
\nopagebreak
\begin{center}
\parbox{15.5cm}{\sl\samepage
Institut f\"ur Physik, Universit\"at Mainz, D-55099 Mainz, Germany$^{16}$}
\end{center}\end{sloppypar}
\vspace{2mm}
\begin{sloppypar}
\noindent
A.~Bonissent,
J.~Carr,
P.~Coyle,
C.~Curtil,
A.~Ealet,
D.~Fouchez,
O.~Leroy,
T.~Kachelhoffer,
P.~Payre,
D.~Rousseau,
A.~Tilquin
\nopagebreak
\begin{center}
\parbox{15.5cm}{\sl\samepage
Centre de Physique des Particules de Marseille, Univ M\'editerran\'ee,
IN$^{2}$P$^{3}$-CNRS, F-13288 Marseille, France}
\end{center}\end{sloppypar}
\vspace{2mm}
\begin{sloppypar}
\noindent
M.~Aleppo,
S.~Gilardoni,
F.~Ragusa
\nopagebreak
\begin{center}
\parbox{15.5cm}{\sl\samepage
Dipartimento di Fisica, Universit\`a di Milano e INFN Sezione di
Milano, I-20133 Milano, Italy.}
\end{center}\end{sloppypar}
\vspace{2mm}
\begin{sloppypar}
\noindent
A.~David,
H.~Dietl,
G.~Ganis,$^{27}$
A.~Heister,
K.~H\"uttmann,
G.~L\"utjens,
C.~Mannert,
W.~M\"anner,
\mbox{H.-G.~Moser},
S.~Schael,
R.~Settles,$^{1}$
H.~Stenzel,
G.~Wolf
\nopagebreak
\begin{center}
\parbox{15.5cm}{\sl\samepage
Max-Planck-Institut f\"ur Physik, Werner-Heisenberg-Institut,
D-80805 M\"unchen, Germany\footnotemark[16]}
\end{center}\end{sloppypar}
\vspace{2mm}
\begin{sloppypar}
\noindent
J.~Boucrot,$^{1}$
O.~Callot,
M.~Davier,
L.~Duflot,
\mbox{J.-F.~Grivaz},
Ph.~Heusse,
A.~Jacholkowska,$^{1}$
L.~Serin,
\mbox{J.-J.~Veillet},
I.~Videau,
J.-B.~de~Vivie~de~R\'egie,$^{28}$
C.~Yuan,
D.~Zerwas
\nopagebreak
\begin{center}
\parbox{15.5cm}{\sl\samepage
Laboratoire de l'Acc\'el\'erateur Lin\'eaire, Universit\'e de Paris-Sud,
IN$^{2}$P$^{3}$-CNRS, F-91898 Orsay Cedex, France}
\end{center}\end{sloppypar}
\vspace{2mm}
\begin{sloppypar}
\noindent
G.~Bagliesi,
T.~Boccali,
G.~Calderini,
V.~Ciulli,
L.~Fo\`a,
A.~Giammanco,
A.~Giassi,
F.~Ligabue,
A.~Messineo,
F.~Palla,$^{1}$
G.~Sanguinetti,
A.~Sciab\`a,
G.~Sguazzoni,
R.~Tenchini,$^{1}$
A.~Venturi,
P.G.~Verdini
\samepage
\begin{center}
\parbox{15.5cm}{\sl\samepage
Dipartimento di Fisica dell'Universit\`a, INFN Sezione di Pisa,
e Scuola Normale Superiore, I-56010 Pisa, Italy}
\end{center}\end{sloppypar}
\vspace{2mm}
\begin{sloppypar}
\noindent
G.A.~Blair,
J.~Coles,
G.~Cowan,
M.G.~Green,
L.T.~Jones,
T.~Medcalf,
J.A.~Strong
\nopagebreak
\begin{center}
\parbox{15.5cm}{\sl\samepage
Department of Physics, Royal Holloway \& Bedford New College,
University of London, Surrey TW20 OEX, United Kingdom$^{10}$}
\end{center}\end{sloppypar}
\vspace{2mm}
\begin{sloppypar}
\noindent
R.W.~Clifft,
T.R.~Edgecock,
P.R.~Norton,
I.R.~Tomalin
\nopagebreak
\begin{center}
\parbox{15.5cm}{\sl\samepage
Particle Physics Dept., Rutherford Appleton Laboratory,
Chilton, Didcot, Oxon OX11 OQX, United Kingdom$^{10}$}
\end{center}\end{sloppypar}
\vspace{2mm}
\begin{sloppypar}
\noindent
\mbox{B.~Bloch-Devaux},$^{1}$
D.~Boumediene,
P.~Colas,
B.~Fabbro,
E.~Lan\c{c}on,
\mbox{M.-C.~Lemaire},
E.~Locci,
P.~Perez,
J.~Rander,
\mbox{J.-F.~Renardy},
A.~Rosowsky,
P.~Seager,$^{13}$
A.~Trabelsi,$^{21}$
B.~Tuchming,
B.~Vallage
\nopagebreak
\begin{center}
\parbox{15.5cm}{\sl\samepage
CEA, DAPNIA/Service de Physique des Particules,
CE-Saclay, F-91191 Gif-sur-Yvette Cedex, France$^{17}$}
\end{center}\end{sloppypar}
\vspace{2mm}
\begin{sloppypar}
\noindent
N.~Konstantinidis,
C.~Loomis,
A.M.~Litke,
G.~Taylor
\nopagebreak
\begin{center}
\parbox{15.5cm}{\sl\samepage
Institute for Particle Physics, University of California at
Santa Cruz, Santa Cruz, CA 95064, USA$^{22}$}
\end{center}\end{sloppypar}
\vspace{2mm}
\begin{sloppypar}
\noindent
C.N.~Booth,
S.~Cartwright,
F.~Combley,
P.N.~Hodgson,
M.~Lehto,
L.F.~Thompson
\nopagebreak
\begin{center}
\parbox{15.5cm}{\sl\samepage
Department of Physics, University of Sheffield, Sheffield S3 7RH,
United Kingdom$^{10}$}
\end{center}\end{sloppypar}
\vspace{2mm}
\begin{sloppypar}
\noindent
K.~Affholderbach,
A.~B\"ohrer,
S.~Brandt,
C.~Grupen,
J.~Hess,
A.~Misiejuk,
G.~Prange,
U.~Sieler
\nopagebreak
\begin{center}
\parbox{15.5cm}{\sl\samepage
Fachbereich Physik, Universit\"at Siegen, D-57068 Siegen, Germany$^{16}$}
\end{center}\end{sloppypar}
\vspace{2mm}
\begin{sloppypar}
\noindent
C.~Borean,
G.~Giannini,
B.~Gobbo
\nopagebreak
\begin{center}
\parbox{15.5cm}{\sl\samepage
Dipartimento di Fisica, Universit\`a di Trieste e INFN Sezione di Trieste,
I-34127 Trieste, Italy}
\end{center}\end{sloppypar}
\vspace{2mm}
\begin{sloppypar}
\noindent
H.~He,
J.~Putz,
J.~Rothberg,
S.~Wasserbaech
\nopagebreak
\begin{center}
\parbox{15.5cm}{\sl\samepage
Experimental Elementary Particle Physics, University of Washington, Seattle,
WA 98195 U.S.A.}
\end{center}\end{sloppypar}
\vspace{2mm}
\begin{sloppypar}
\noindent
S.R.~Armstrong,
K.~Cranmer,
P.~Elmer,
D.P.S.~Ferguson,
Y.~Gao,$^{29}$
S.~Gonz\'{a}lez,
O.J.~Hayes,
H.~Hu,
S.~Jin,
J.~Kile,
P.A.~McNamara III,
J.~Nielsen,
W.~Orejudos,
Y.B.~Pan,
Y.~Saadi,
I.J.~Scott,
\mbox{J.H.~von~Wimmersperg-Toeller}, 
J.~Walsh,
W.~Wiedenmann,
J.~Wu,
Sau~Lan~Wu,
X.~Wu,
G.~Zobernig
\nopagebreak
\begin{center}
\parbox{15.5cm}{\sl\samepage
Department of Physics, University of Wisconsin, Madison, WI 53706,
USA$^{11}$}
\end{center}\end{sloppypar}
}
\footnotetext[1]{Also at CERN, 1211 Geneva 23, Switzerland.}
\footnotetext[2]{Now at Universit\'e de Lausanne, 1015 Lausanne, Switzerland.}
\footnotetext[3]{Also at Dipartimento di Fisica di Catania and INFN Sezione di
 Catania, 95129 Catania, Italy.}
\footnotetext[4]{Deceased.}
\footnotetext[5]{Also Istituto di Cosmo-Geofisica del C.N.R., Torino,
Italy.}
\footnotetext[6]{Supported by the Commission of the European Communities,
contract ERBFMBICT982894.}
\footnotetext[7]{Supported by CICYT, Spain.}
\footnotetext[8]{Supported by the National Science Foundation of China.}
\footnotetext[9]{Supported by the Danish Natural Science Research Council.}
\footnotetext[10]{Supported by the UK Particle Physics and Astronomy Research
Council.}
\footnotetext[11]{Supported by the US Department of Energy, grant
DE-FG0295-ER40896.}
\footnotetext[12]{Now at Departement de Physique Corpusculaire, Universit\'e de
Gen\`eve, 1211 Gen\`eve 4, Switzerland.}
\footnotetext[13]{Supported by the Commission of the European Communities,
contract ERBFMBICT982874.}
\footnotetext[14]{Also at Rutherford Appleton Laboratory, Chilton, Didcot, UK.}
\footnotetext[15]{Permanent address: Universitat de Barcelona, 08208 Barcelona,
Spain.}
\footnotetext[16]{Supported by the Bundesministerium f\"ur Bildung,
Wissenschaft, Forschung und Technologie, Germany.}
\footnotetext[17]{Supported by the Direction des Sciences de la
Mati\`ere, C.E.A.}
\footnotetext[18]{Supported by the Austrian Ministry for Science and Transport.}
\footnotetext[19]{Now at SAP AG, 69185 Walldorf, Germany}
\footnotetext[20]{Now at Harvard University, Cambridge, MA 02138, U.S.A.}
\footnotetext[21]{Now at D\'epartement de Physique, Facult\'e des Sciences de Tunis, 1060 Le Belv\'ed\`ere, Tunisia.}
\footnotetext[22]{Supported by the US Department of Energy,
grant DE-FG03-92ER40689.}
\footnotetext[23]{Now at Department of Physics, Ohio State University, Columbus, OH 43210-1106, U.S.A.}
\footnotetext[24]{Also at Dipartimento di Fisica e Tecnologie Relative, Universit\`a di Palermo, Palermo, Italy.}
\footnotetext[25]{Now at CERN, 1211 Geneva 23, Switzerland.}
\footnotetext[26]{Now at ISN, Institut des Sciences Nucl\'eaires, 53 Av. des Martyrs, 38026 Grenoble, France.}
\footnotetext[27]{Now at INFN Sezione di Roma II, Dipartimento di Fisica, Universit\`a di Roma Tor Vergata, 00133 Roma, Italy.}
\footnotetext[28]{Now at Centre de Physique des Particules de Marseille,Univ M\'editerran\'ee, F-13288 Marseille, France.}
\footnotetext[29]{Also at Department of Physics, Tsinghua University, Beijing, The People's Republic of China.}
\footnotetext[30]{Also at SLAC, Stanford, CA 94309, U.S.A.}
%
\setlength{\parskip}{\saveparskip}
\setlength{\textheight}{\savetextheight}
\setlength{\topmargin}{\savetopmargin}
\setlength{\textwidth}{\savetextwidth}
\setlength{\oddsidemargin}{\saveoddsidemargin}
\setlength{\topsep}{\savetopsep}
\normalsize
\newpage
\pagestyle{plain}
\setcounter{page}{1}

\newpage
\pagenumbering{arabic}

\parskip 0.25cm plus 0.10cm
\section{Introduction}
In this letter the results of searches for
sleptons (\PSl), squarks (\PSq), charginos (\Pcha, \Pchap) and 
neutralinos (\PChiz{i}) are reported, obtained 
with the data collected by the ALEPH detector at LEP during 
1999 (for squark and slepton searches) and 1998-1999 (for chargino and 
neutralino searches), 
at centre-of-mass energies ranging from 188.6 to 201.6\,GeV. 
Energies and integrated luminosities of 
the analysed data samples are given in Table~\ref{Tab:stat}.
Results of slepton and squark searches 
with the 1998 data sample have already been reported in 
Ref.~\cite{scalars98}.
As in Refs.~\cite{scalars98,chaneu96,chaneu97}, 
the theoretical framework is the Minimal Supersymmetric extension 
of the Standard Model (MSSM), 
with R-parity conservation and the assumption that the 
lightest neutralino is the Lightest Supersymmetric Particle (LSP).
The notations and conventions described in Ref.~\cite{chaneu96} are used 
for the MSSM parameters.
\begin{table}[ht]
\caption{\small
  Definition of the analysed data samples.}
\label{Tab:stat}
 \begin{center}
 \begin{tabular}{|l||c||c|c|c|c|}
  \hline
         & 1998 & \multicolumn{4}{|c|}{1999 } \\
\hline
$\rts$ (GeV)       & 188.6 & 191.6 & 195.5 & 199.5 & 201.6 \\
$\Lumi$ ($\pbinv$) & 174.2 & 28.9 & 79.8 & 86.2 & 42.0 \\
\hline
\end{tabular}
\end{center}
\end{table}

The final state topologies addressed by the searches are 
summarized in Table~2 together with the related signal processes.
For a given final state topology, various selection criteria are applied 
which depend mainly on the mass difference \dm\ between the produced particle 
and the LSP. The selection algorithms are basically the same as in previous 
publications~\cite{scalars98,chaneu96,chaneu97} but,
in order to cope with the increased centre-of-mass energy and  
with the larger size of the new data samples,
the positions of the cuts were re-optimized so as to give the 
lowest expected upper limit on the number of produced signal events 
in the case of absence of signal.

For the interpretation of the results in the MSSM, the unification relation 
among the gaugino supersymmetry breaking mass terms,
$M_1\!=\!{5\over 3}M_2\tan^2\theta_W$, is assumed.
The region where $M_2 \gg|\mu|$ is referred to as the {\it higgsino region},
and the region where $\vert\mu\vert\gg M_2$ as the {\it gaugino region}.
Unless otherwise specified, all supersymmetric particle 
masses and couplings are calculated at tree level. 
For charginos and neutralinos, they are entirely determined by $M_2$, 
the Higgs mass term $\mu$, 
and the ratio of the vacuum expectation values of the 
two Higgs doublets, commonly indicated with \tanb. 
When relevant, the sfermion masses are 
calculated from the renormalization group equations assuming a common 
supersymmetry breaking mass term $m_0$ for all sleptons and squarks at 
the grand-unification scale. 
The results in the gaugino sector, in particular the LSP mass lower limit, 
are derived under the assumption of flavour-independent leptonic branching
ratios. 
The possible impact on these results of a large mixing in the stau sector is 
therefore not considered in this letter. 

\renewcommand{\arraystretch}{1.1}
\begin{table}[ht]
\caption{\small 
  Topologies studied in the different searches; only 
  the main decay chains contributing to the different topologies 
  are indicated for neutralinos.}
\label{topologies}
 \begin{center}
 \begin{tabular}{|l|l|l|}
  \hline
Production  & Decay mode & Topology  \\
\hline
\hline
 $\PSl \bar{\PSl}$ & $\PSl \to \ell \PChiz{1}$ & Acoplanar leptons   \\
 $\PSe_{\rm{L(R)}} \bar{\PSe}_{\rm{R(L)}}$ & $\PSe \to \rm{e} \PChiz{1}$ &
 Single electron  (small $m_{\PSe_{\rm{R}}}\!-\!m_{\PChiz{1}}$) \\
\hline\hline
\rule{-4pt}{14pt} $\PSq\bar{\PSq}$ &
 $\PSq \to \rm{q} \PChiz{1}$ & Acoplanar jets \\
 $\PSt\bar{\PSt}$ &
 $\PSt \to \rm{c} \PChiz{1}$ & Acoplanar jets \\
$\PSb\bar{\PSb}$ &
 $\PSb \to \rm{b} \PChiz{1}$ & Acoplanar b-jets \\

 $\PSt \bar{\PSt}$ & $\PSt \to \rm{b} \ell \PSnu $ &
Acoplanar jets plus leptons \\
\hline\hline
$\Pchi^+\Pchi^-$ & $\Pcha\!\to\!\mathrm{q \bar{q}^\prime}\PChiz{1}$ & 4 jets + $\not{\!\!E}$  \\
           & $\Pcha\!\to\!\ell^\pm \nu\PChiz{1}$ & Acoplanar leptons  \\
           & mixed & 2 jets + lepton + $\not{\!\!E}$  \\
\hline\hline
\neui\neuj         & $\neuu\neuj\!\to\!\mathrm{q \bar{q}}\neuu$ & 
                            \raisebox{-3mm}[0mm][-3mm]{Acoplanar jets} \\
                   & $\neui\neuj\!\to\!\nu\bar{\nu}\neuu 
                      \mathrm{q \bar{q}}\neuu,\ldots$ & \\
$j\!\geq\!i, j\!\neq\!1$ & $\neuu\neuj\!\to\!\ell^+\ell^-\neuu$ 
                            & \raisebox{-3mm}[0mm][-3mm]{Acoplanar leptons} \\
                         & $\neui\neuj\to\!\nu\bar{\nu}\neuu \ell^+\ell^-\neuu,\ldots$ 
                                                        &  \\
\hline
\end{tabular}
\end{center}

\end{table}
\renewcommand{\arraystretch}{1.0}

The results of Higgs boson searches~\cite{higgs202} are used to 
further constrain the MSSM 
parameter space as discussed in Ref.~\cite{chaneu97}. 
The inclusion of the Higgs boson searches requires that 
the pseudo-scalar neutral Higgs boson mass, $\mA$, and the 
trilinear coupling in the stop sector, $A_{\mathrm t}$, be 
included in the analysis.  
Masses and mixing angles in the Higgs sector are obtained with the 
two-loop level calculations of Ref.~\cite{CW}, as implemented in 
Ref.~\cite{PaJ}.

As in Ref.~\cite{chaneu97}, the results are also interpreted in the framework 
of a highly constrained MSSM version known as minimal supergravity. 
Masses and couplings are calculated in terms of five parameters: 
the mass term $m_0$ common to all scalars 
(Higgs bosons, squarks and sleptons), 
the common supersymmetry-breaking gaugino mass term $m_{1/2}$, and a common 
trilinear coupling $\Azero$ (all defined at the grand unification scale),
$\tanb$ and the sign of $\mu$.
To solve the appropriate set of renormalization group equations, 
the latest version of the ISAJET package~\cite{isajet} 
is used. For this analysis, the one-loop radiative corrections to chargino 
and neutralino masses \cite{efgos} are included. 

This letter is organized as follows. 
The modifications to the selection algorithms and the results of the searches 
are described in Section~\ref{sec:updates}. The interpretation of these 
results in the theoretical frameworks mentioned above is presented
in Section~\ref{sec:interpretation}, with a special focus on the lower limit 
on the LSP mass. The conclusions are given in Section~\ref{sec:conclusions}. 

A thorough description of the ALEPH detector and of its performance, 
as well as of the standard reconstruction and analysis algorithms,
can be found in Refs.~\cite{AlephDetector,AlephPerformances}. 
Only a brief summary is given here. 
Charged particle tracking, down to 16$^\circ$ from the beam axis, is obtained 
by means of a silicon vertex detector, a cylindrical drift chamber, and a 
large time projection chamber, all immersed in a 1.5~T axial magnetic field 
provided by a superconducting solenoidal coil. 
Hermetic calorimetric coverage, down to polar angles of 34~mrad, 
is achieved by means of a highly granular electromagnetic calorimeter, by
dedicated low angle luminosity monitors, and by the iron return yoke 
instrumented to act as a hadron calorimeter. The latter is  
supplemented with external muon chambers. 
The information from all these detectors is combined in an energy-flow 
algorithm~\cite{AlephPerformances} 
which provides a list of charged 
particles (electrons, muons, charged hadrons), photons and neutral 
hadrons, used also 
to determine global quantities such as total energy or missing momentum. 
The resolution achieved on the total visible energy is 
$(0.6\sqrt{E}+0.6)$\,GeV ($E$ in\,GeV).

\section{Update of sparticle searches}
\label{sec:updates}

The simulation of the signal production, which includes a detailed treatment of 
cascade decays as well as of initial and final state radiation, was performed 
with {\tt SUSYGEN}~\cite{SUSYGEN}.
For the determination of the selection efficiencies,  
a fast detector simulation, 
cross-checked with several fully simulated samples, was used.

Standard model processes were simulated with
{\tt BHWIDE}~\cite{BHWIDE} for Bhabha production, 
{\tt KORALZ}~\cite{KORALZ} for $\mu^+\mu^-$ and $\tau^+\tau^-$ production, 
{\tt PHOT02}~\cite{PHOT02} for  \ggll\ and tagged \gghh, 
{\tt PHOJET}~\cite{PHOJET} for untagged \gghh, 
{\tt KORALW}~\cite{KORALW} for \ww, 
{\tt Grace4F}~\cite{GRC4F} for \wen, 
a private generator~\cite{ZNNBAR} for $\znn$ events, 
and {\tt PYTHIA}~\cite{PYTHIA} for all other processes.
The integrated luminosity of the simulated samples corresponds 
to at least 25 times the integrated luminosity of the data sample, 
except for \gaga\ processes for which the statistics are at least a factor of 
three larger than that of the data sample. 
All standard model background samples were processed through the full detector 
simulation. 

Background subtraction was generally performed when optimizing the 
selection algorithms, except in searches for squarks   
and for hadronic and mixed final states arising from chargino production, 
where the expected gain is marginal.

\subsection{Update of sfermion searches}

\label{sec:updatesfer}
The final states studied (Table~2) are those
arising from squark and slepton pair-production followed 
by the decays 
$\PSt \to \rm{c}\PChiz{1}$, $\PSt \to\rm{b} \ell \PSnu$,
$\PSb \to \rm{b} \PChiz{1}$, $\PSq \to \rm{q} \PChiz{1}$,
$\PSe \to \rm{e} \PChiz{1}$, $\PSmu \to \mu \PChiz{1}$ and $\PStau \to \tau \PChiz{1}$.
 Events with acoplanar jets and acoplanar jets plus two leptons
 are signatures for squark production.
 Events with acoplanar lepton pairs or with single electrons  
 are expected from slepton production.
All these final states are characterized by missing energy.
To reach the best sensitivity to the  expected signal for a large 
range of $\dm$ values, two different selection procedures are 
employed~\cite{scalars98,squarks97}, 
which specifically address the small and the large $\dm$ cases.
Systematic uncertainties on selection efficiencies 
 and background estimations remain at the few percent level,  
 as in Ref.~\cite{scalars98}.

The numbers of events selected in the 1999 data by the sfermion searches 
are reported in Table~3, together with the results obtained with the 1998 
data sample. In general, agreement is observed between 
numbers of candidate events and expectations from standard processes. 
However, a slight excess is observed in the acoplanar tau search; 
the probability 
for an upward statistical fluctuation of the expected background is  
1.6\% for the 1999 sample alone, and 1.2\% when the 1998 data are included. 
The events selected by the $\PSb$ searches are also  
selected by the $\PSt \to \rm{c} \PChiz{1}$ analyses. 


\subsection{Update of gaugino searches}

\label{sec:updategaug}

\subsubsection*{Chargino searches}

Searches are performed in all possible chargino-pair 
decay topologies such as four-jets (4J),
hadrons plus electron or muon (2J$\ell$),
hadrons plus tau (2J$\tau$) and acoplanar lepton pairs (A$\ell$), 
as described in Refs.\cite{chaneu96,chaneu97}.
The selection efficiencies, the background contaminations and 
the main systematic uncertainties, related to the simulation 
of the energy-flow reconstruction and of the lepton identification, 
are similar to those reported for the analyses at lower energies. 
For each set of mass difference and leptonic branching ratio, 
the optimal combination of selections is determined as explained 
in Ref.~\cite{chaneu97}.

\subsubsection*{Neutralino searches}

In the case of large sfermion masses (large \mzero), the neutralino searches
are applied only in the higgsino region where the final state is characterized 
by acoplanar jets. The sensitivities of the two analyses, dedicated to small 
and large \dm, 
are significantly improved by subtracting the expected four-fermion 
background when re-optimizing the cuts.
The selection efficiencies remain comparable to those of the analyses applied 
to the 183\,GeV data sample~\cite{chaneu97}. 

For small sfermion masses, the leptonic 
branching ratios are enhanced.
A new search was developed to address acoplanar electron or 
muon final states. 
The selection consists of the preselection of the selectron and smuon searches 
supplemented with a final sliding cut on the sum of the two lepton energies.
The exact location of this cut depends on the mass difference $\dm$ 
between the \PChiz{j}\  and the \PChiz{1}.
Selection efficiencies of 70\% are reached for 
a large range of $\dm$ values, with an irreducible background of  
0.37~pb, mainly due to $\wwll$ events.
The systematic uncertainties associated to this new selection are similar 
to those quoted for the selectron and smuon searches~\cite{scalars98}. 

\subsubsection*{Results}

The numbers of events observed and expected from standard processes for the 
chargino and neutralino selections are reported in Table~3. 
The numbers of candidate events are in agreement with the expected background 
for all the selections.

\renewcommand{\arraystretch}{1.2}

\begin{table}[ht]
\caption{\small 
Numbers of candidate events observed in the data ($N_{\rm{cand}}$) and 
background events expected from standard model processes ($N_{\rm{bkg}}$).}
\label{Tab:candsfe}
 \begin{center}
 \begin{tabular}{|c|c||c|c||c|c|}
  \hline
 Process & Comment &   
  \multicolumn{2}{|c||}{1998 data} & \multicolumn{2}{|c|}{ 1999 data} \\
\cline{3-6}
& & $N_{\mathrm{cand}}$  & $N_{\mathrm{bkg}}$ & $N_{\mathrm{cand}}$  & $N_{\mathrm{bkg}}$ \\
\hline
 \PSe\PSe     & & 33 & 32.8  & 42    & 48.1  \\
\hline
 \PSmu\PSmu   & & 28 & 29.6  & 39    & 43.4  \\
\hline
 \PStau\PStau & & 26 & 21.5  & 46    & 32.7  \\
\hline
$\PSe_{\mathrm{L(R)}} \bar{\PSe}_{\mathrm{R(L)}}$ 
              &     &  8 & 13.8  & 22    & 22.4  \\
\hline\hline
$\PSt \to \rm{c} \PChiz{1}$ 
              & small \dm\  & 3 & 5.5  &  2    & 2.4   \\
              & high \dm\ & 5 & 4.0  &  8    & 7.3   \\
\hline
$\PSb \to \rm{b} \PChiz{1}$ 
              & small \dm\  & 3 & 3.3  &  1    & 2.2   \\
              & high \dm\ & 0 & 0.9 &  1    & 0.7   \\
\hline
$\PSt \to \rm{b} \ell \PSnu$ 
              & small \dm\  & 0 & 1.9 &  3    & 2.6   \\
              & high \dm\ & 2 & 0.4 &  2    & 1.4   \\
\hline
\hline
Chargino & $\mathrm{W}^*$ branching ratios &
          10 & 8.3  & 9    & 12.7  \\
         & any branching ratios &
          25 & 23.0 & 24   & 33.9  \\
\hline\hline
Neutralino & Acop.~Jets ($\dm\!<\!40\ \Gcsq$) &
           4 & 3.0 & 6    & 4.5  \\
           & Acop. Leptons &
           59 & 60.0 & 76 & 79.8  \\
\hline
\end{tabular}
\end{center}

\end{table}
\renewcommand{\arraystretch}{1.}

\section{Interpretation of the results}

\label{sec:interpretation}

 As no significant excess of candidate events is observed, 
 upper limits on the production cross sections are set  
 for the processes searched for. 
 Each candidate event contributes to a limited range of $\dm$.
 The systematic uncertainties on the selection  efficiencies  are included  
 following the method described in Ref.~\cite{syse}.
 When background subtraction is performed, the prescription of 
 Ref.~\cite{PDG96}  is adopted, with the number of subtracted background 
 events conservatively reduced by one standard deviation of its systematic 
 uncertainty.
 Finally, the constraints presented in this section are derived by combining 
 the searches presented here with those reported 
 in Refs.~\cite{scalars98,chaneu96,chaneu97}, and are at 95\% 
 confidence level.

\subsection{Slepton production}

\begin{figure}[p]
\begin{center}
 \mbox{\epsfig{file=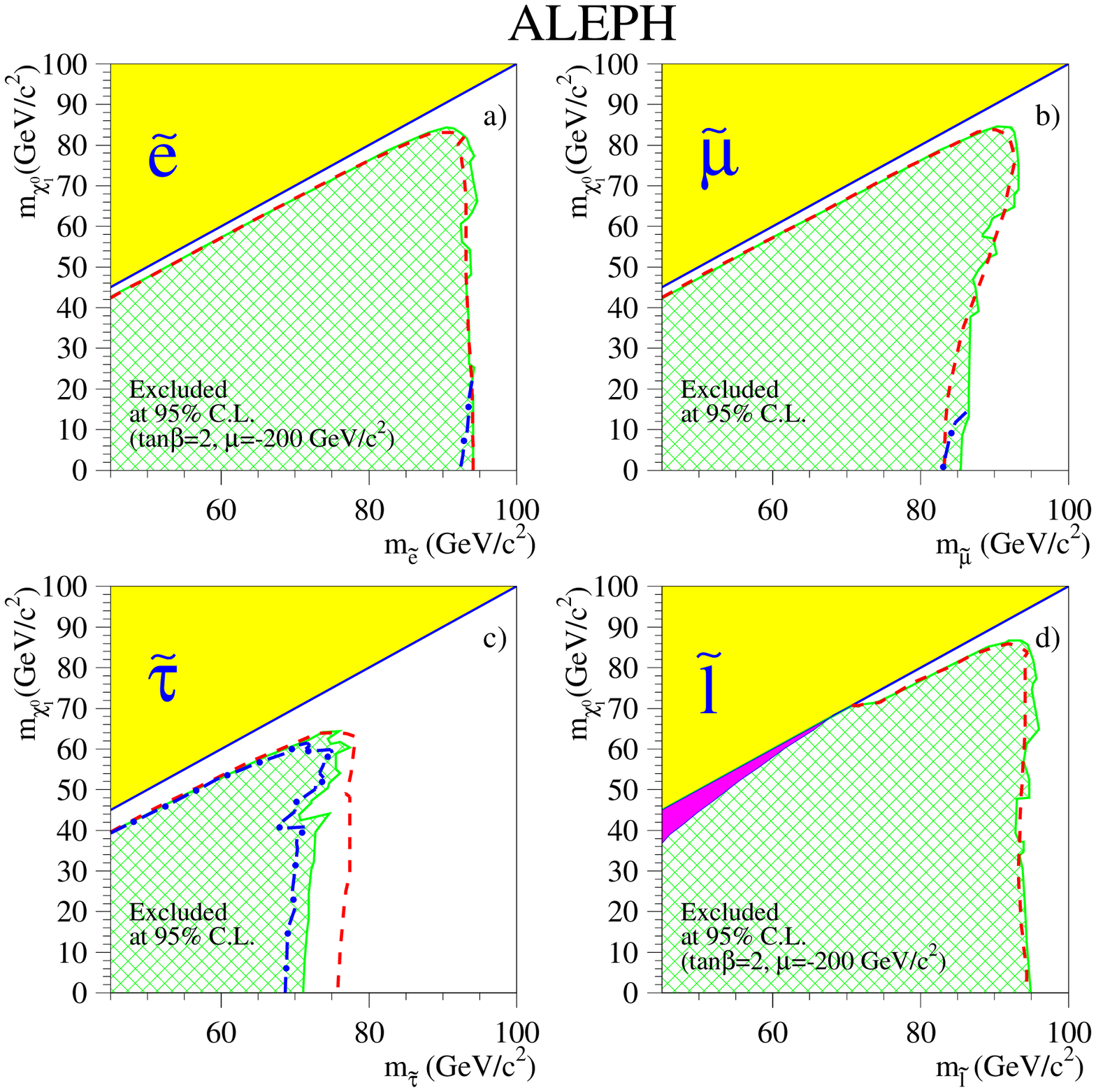,height=16cm}}
\end{center}
 \caption{\label{slep_lim} \small
 Excluded regions at 95\% C.L. in the $m_{\tilde{\ell}_{\rm{R}}}$
 vs $m_{\PChiz{1}}$ plane from slepton searches assuming 
 BR$(\rm{\tilde{ \ell}} \rightarrow \ell \PChiz{1}) = 100\%$ 
 (cross-hatched regions); the dashed curves show the expected 
 exclusion under the same assumptions. 
 The dot-dashed curves in {\it a)} and {\it b)} show the 
 effect of cascade decays for $\tan \beta = 2$ and $\mu = -200$~\Gcsq, 
 assuming zero efficiency for those decays.
 The dot-dashed curve in {\it c)} shows the  limit in the case of minimal 
 $\tilde{\tau}_1 \tilde{\tau}_1$Z coupling. 
 In {\it d)}, the effect of cascade decays is included.
 The dark shaded region is not accessible because the common scalar mass
 at the GUT scale becomes unphysical.
 In all the plots, the light shaded region corresponds
 to the $m_{\PSl}<m_{\PChiz{1}}$ disallowed region. 
 }
\end{figure}

The cross section upper limits, translated into excluded regions in 
the $(m_{{\PSl}_{\mathrm R}},\Mchi)$ plane, are shown in Figs.~\ref{slep_lim}a-c 
for $\PSl^+_{\mathrm R}\PSl^-_{\mathrm R}$ production and
a 100\% $\PSl_{\mathrm R}\!\to\!\ell \PChiz{1}$ branching ratio. 
For selectrons, it is 
assumed that $\tan \beta = 2$ and $\mu = -200\,\Gcsq$. 
For $\PSe$ and $\PSmu$, the impact of cascade decays such as 
$\PSl_{\mathrm R}\!\to\!\ell \PChiz{j}$ $(j\!>\!1)$ 
is illustrated by the excluded region obtained with a vanishing selection 
efficiency for final states deriving from those decays. 
For $\dm\!>\!10\,\Gcsq$, the lower limits 
on $m_{{\PSe}_{\mathrm R}}$, $m_{{\PSmu}_{\mathrm R}}$ and $m_{{\PStau}_{\mathrm R}}$ 
are 92, 85 and 70\,\Gcsq, respectively. 
For staus, the exclusion obtained for the mixing angle corresponding to the 
minimal cross section is also shown. In such a case, the lower limit 
on the mass is 68\,$\Gcsq$ for $\dm\!>\!10\,\Gcsq$.

The results of the searches for acoplanar leptons are combined with those 
of the search for events with single electrons, 
using the slepton masses determined from the GUT relations,
 and with the additional assumption of no mixing in 
the stau sector. The region of the plane $(m_{\PSl_{\mathrm R}},\Mchi)$ 
excluded for $\tan\beta=2$ and $\mu=-200$\,\Gcsq\ is 
shown in Fig.~\ref{slep_lim}d. 
 The loss of sensitivity of the $\PSe_{\mathrm R} \bar{\PSe}_{\mathrm R}$
 search for $\dm\!<\!3\,\Gcsq$ is recovered for $m_{\PSl}<70\,\Gcsq$ 
by the search for $\PSe_{\mathrm R}^{\pm} \PSe_{\mathrm L}^{\mp}$. 
The effect of an assumed null efficiency for cascade decays at small $\Mchi$ 
is compensated by the contribution of  the $\PSl_{\mathrm L} \PSl_{\mathrm L}$ 
production.

\subsection{Squark production}

\begin{figure}[p]
\begin{center}
 \mbox{\epsfig{file=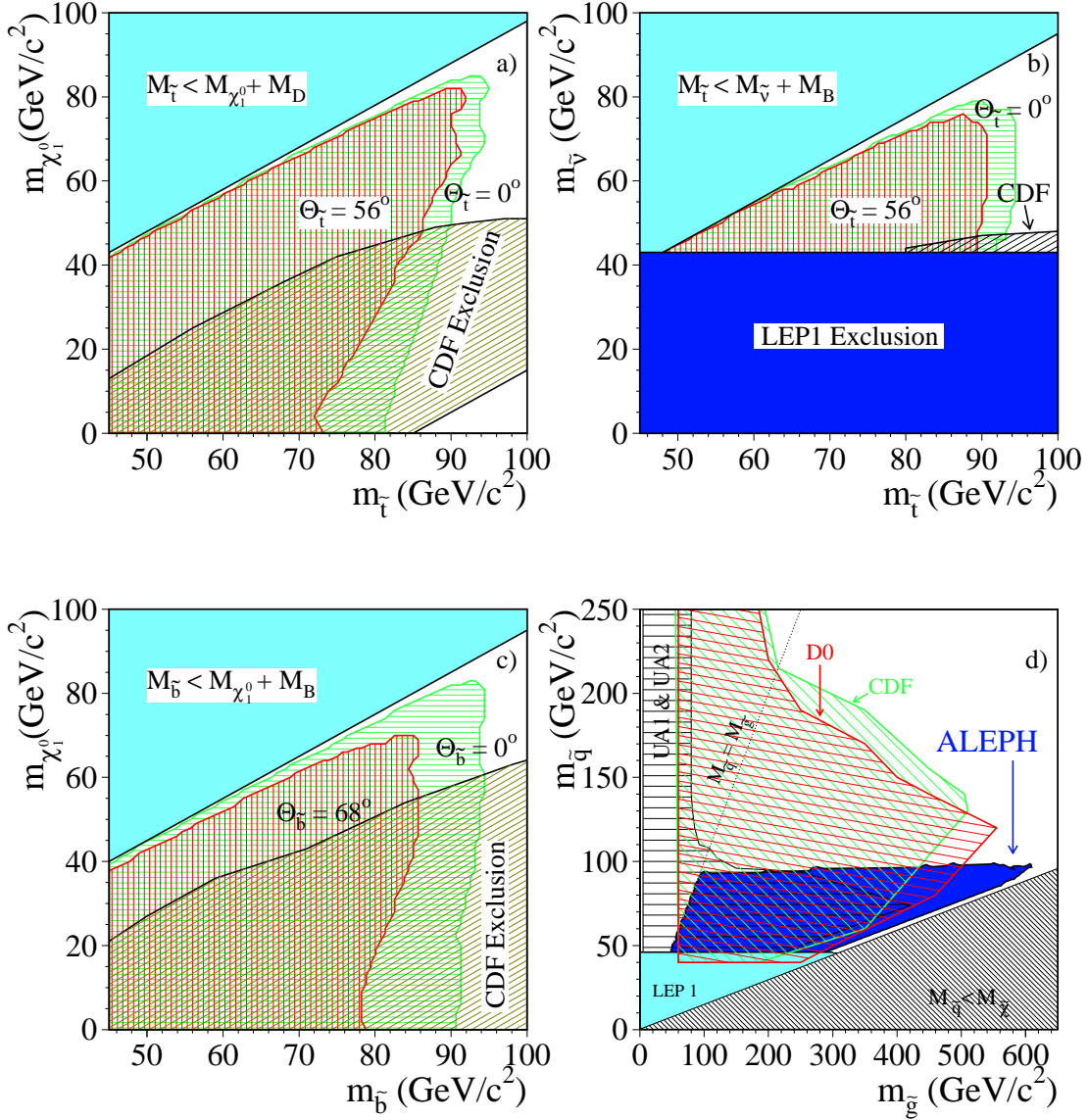,height=18cm}}
\end{center}
\caption{ \small 
 Excluded regions at 95\% C.L. from squark searches: {\it a)}  
 in the ($m_{\neuu}$,$m_{\PSt}$) plane assuming a dominant  
 $\PSt \rightarrow \mathrm{c}\neuu$ decay;
 {\it b)} in the ($m_{\tilde{\nu}}$,$m_{\PSt}$) plane for
a dominant $\PSt \rightarrow \mathrm{b}\ell\tilde{\nu}$
decay and equal branching fractions into $\mathrm{e}$, $\mu$, and $\tau$;
{\it c)} in the ($m_{\neuu}$,$m_{\PSb}$) plane for a dominant
 $\PSb \rightarrow\mathrm{b}\neuu$ decay; 
 {\it d)} in the plane $(m_{\PSgl},m_{\PSq})$ in the case of 
 five mass degenerate $\PSq$  flavours, {\mbox{$\tan\beta = 4$}} and 
$\mu$ = $-400\,\Gcsq$. 
 The hadron collider results are also shown for comparison. 
\label{squarks}}
\end{figure}

Under the assumption of a dominant 
$\PSt\to \mathrm{c} \neuu$ decay, 
the regions excluded by stop searches in the plane $(m_{\PSt},\Mchi)$ 
 are shown in Fig.~\ref{squarks}a
 for two values of the $\PSt$ mixing angle $\thetat$, $0^\circ$ 
 and $56^\circ$, 
 corresponding to maximal and minimal cross section, respectively.
 For $\dm$ in the range from 6 to 40 $\Gcsq$, {\it i.e.}, a region not 
 accessible to the Tevatron searches, 
 the lower limit on $m_{\PSt}$ is  83~$\Gcsq$, independent of $\thetat$.
 In the case of a dominant $\mathrm \PSt\!\to\! {\rm b}\ell\widetilde{\nu}$ 
 decay, the excluded region in the plane $(m_{\PSt},m_{\PSnu})$
 is shown in Fig.~\ref{squarks}b, where equal branching ratios for 
 $\ell=\mathrm{e},\ \mu$ and $\tau$ are assumed.  
 For $\dm\!>\!10\,\Gcsq$ and with the LEP~1 lower limit on the 
 sneutrino mass 
 of 43~$\Gcsq$~(obtained for three mass degenerate $\tilde{\nu}$'s), 
 the $\thetat$--independent lower limit on $m_{\PSt}$ is 88~$\Gcsq$.
 The regions excluded by sbottom searches in the plane $(m_{\PSb},\Mchi)$ 
 under the assumption of a dominant $\PSb\!\to\!{\rm b}\PChiz{1}$ decay
 are shown in Fig.~\ref{squarks}c for two values of the $\PSb$ mixing angle 
 $\thetab$, corresponding to minimal ($\thetab\!=\!68^{\circ}$) and maximal 
 ($\thetab\!=\!0^{\circ}$) production cross section. In the latter case, 
 and for $\dm\!>\!10\,\Gcsq$, a lower limit of 91~$\Gcsq$ is set on $m_{\PSb}$.

 In Fig.~\ref{squarks}d, 
 the negative outcome of the searches for acoplanar jets is translated into 
 exclusion domains in the plane $(m_{\tilde{\mathrm{g}}},m_{\PSq})$
  for mass--degenerate squarks (except the two stop particles) and with  
 unification of the gluino and weak gaugino masses. Here 
 $\tan\beta\!=\!4$ and $\mu\!=\!-400$\,\Gcsq\ are chosen, as in 
 Refs.~\cite{tev_cdf,tevatron2}. 

 The results obtained at the Tevatron~\cite{tev_cdf, tevatron2} are also shown 
 in Fig.~\ref{squarks}.  The sensitivity of 
  the squark searches presented in this letter extends to smaller $\dm$ values 
  than those tested at hadron colliders.

\subsection{Gaugino production}

\begin{figure}[p]
\begin{center}
 \mbox{\epsfig{file=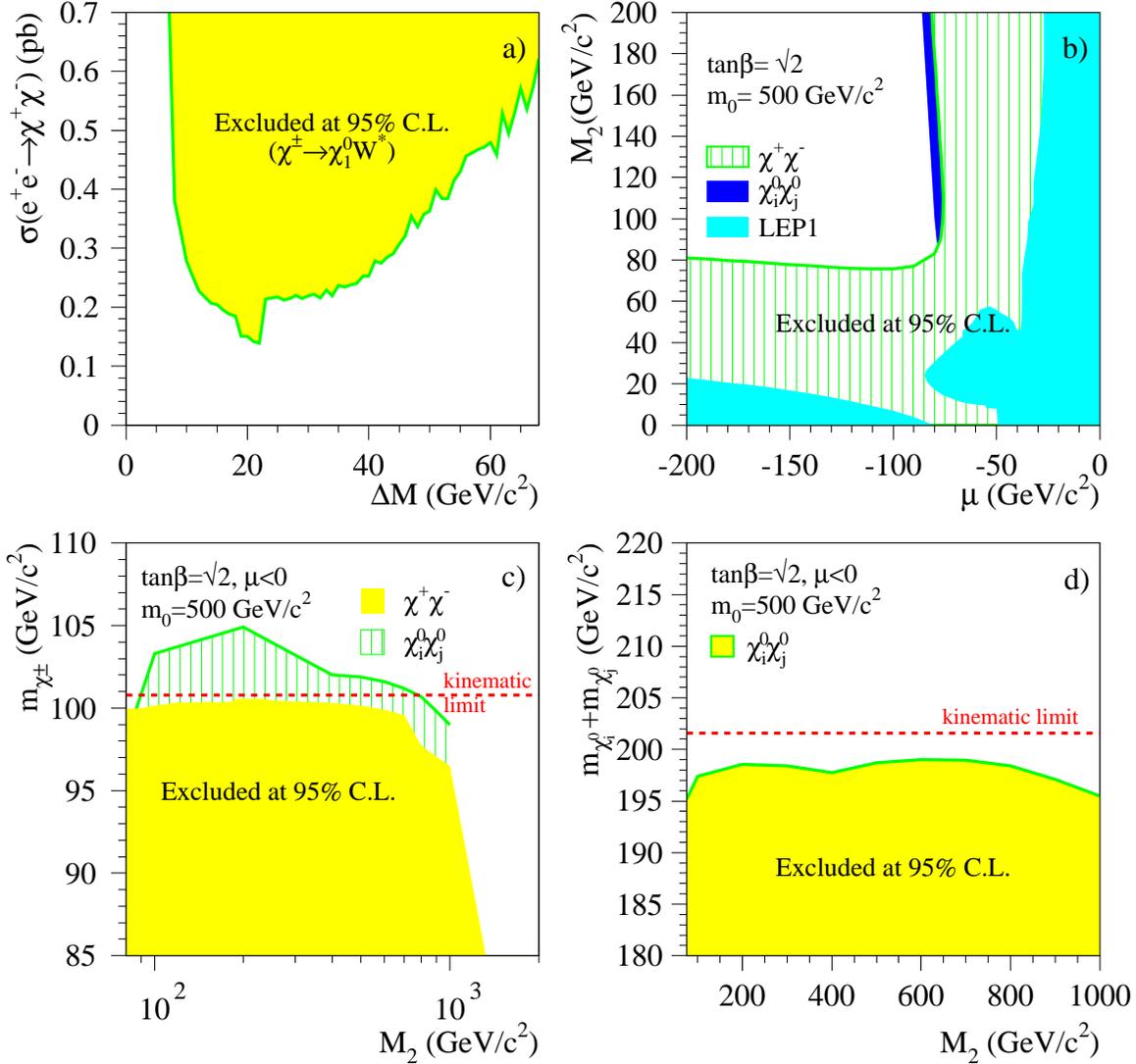,height=16cm}}
\end{center}
\caption{\small Gaugino searches: {\it a)} upper limit 
on the chargino pair production cross section for chargino masses
 close to the kinematic limit at $\sqrt{s}=201.6$\,GeV; {\it b)} 
excluded domains in the $(\mu, M_2)$ plane of the MSSM; 
{\it c)} chargino 
mass lower limit in the higgsino region; 
{\it d)} lower limit on the sum of the masses of two 
 neutralinos produced with the largest cross section ($\PChiz{1}\PChiz{2}$ 
 for large $M_2$ and mainly $\PChiz{1}\PChiz{3}$ for small $M_2$). 
In b, c, and d, it is assumed that $m_0=500$\,GeV$/c^2$, $\mu < 0$ and 
$\tan\beta=\sqrt{2}$. 
}
\label{fig:gaug}
\end{figure}

In Fig.~\ref{fig:gaug}a, the upper limit on the production cross 
section is displayed for chargino pairs with masses close to the 
kinematic limit at 
$\sqrt{s}=201.6$\,GeV. It is assumed that charginos decay through the process
$\Pcha\!\to\!\PChiz{1}{\mathrm W}^{\pm*}$. 
In Figs.~\ref{fig:gaug}b-d, the gaugino cross section upper limits are 
translated, for $\tanb=\sqrt 2$ and large $m_0$, 
into exclusion domains in the $(\mu,M_2)$ plane of the MSSM,  
into lower limits on the chargino mass, and into lower limits on 
$m_{\PChiz{1}}\!+\!m_{\PChiz{2}}$. 
The kinematic limit is closely approached for both 
chargino and associated neutralino production, except in the deep higgsino 
region in which $\dm$, and hence the selection efficiencies, are small. 
For $\tanb=\sqrt 2$ and negative 
$\mu$, the indirect chargino mass limit derived from the neutralino searches 
extends beyond the chargino kinematic limit by up to 5\,\Gcsq.

\subsection{Lower limit on the LSP mass}

\begin{figure}[p]
 \begin{center}
 \mbox{\epsfig{file=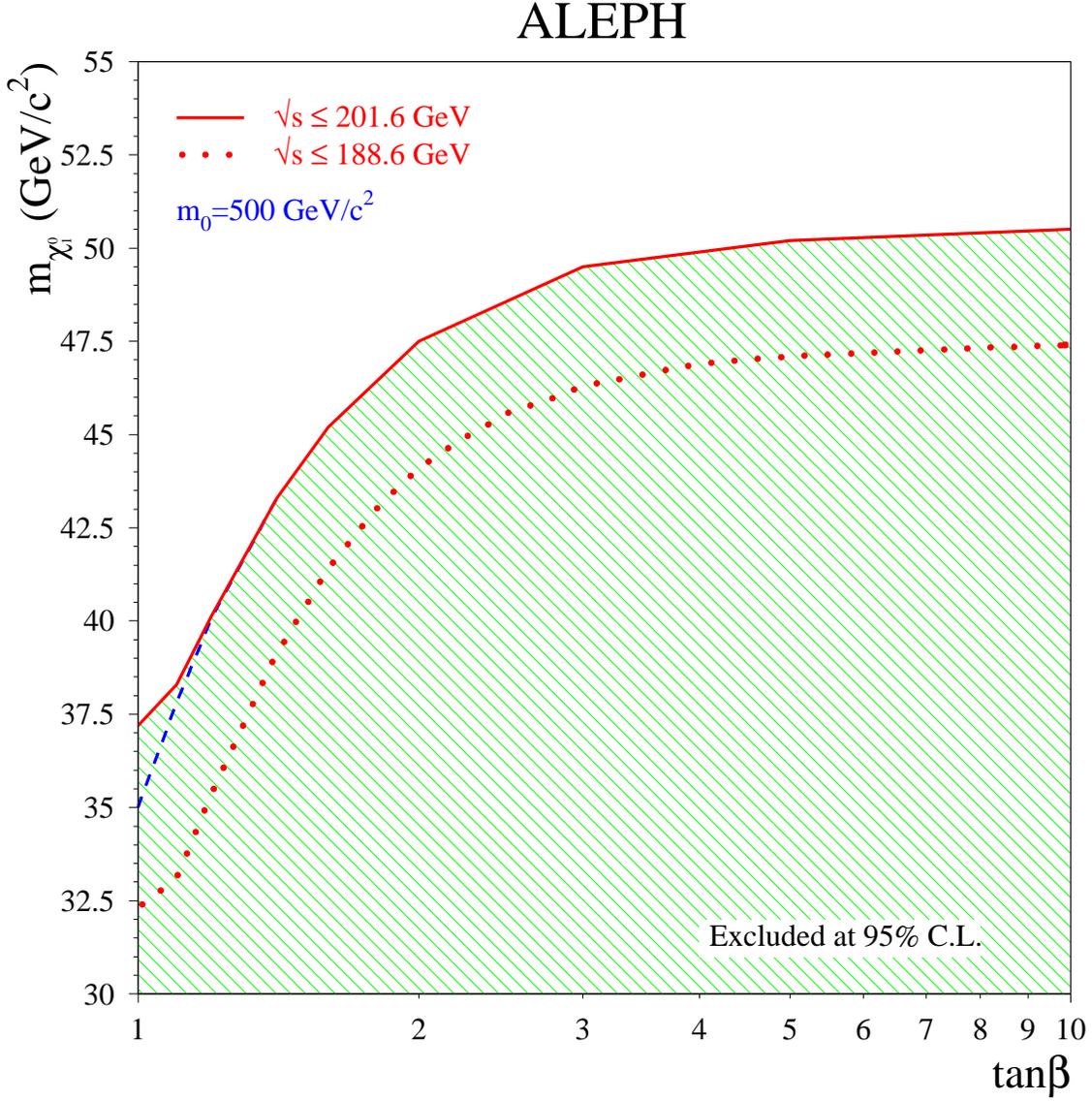,height=16cm}}
 \end{center}
 \caption{\small The 95\% C.L. lower limit on the mass of the lightest neutralino 
 as a function of $\tanb$ for 
 $m_0=500$\,GeV$/c^2$. The dashed line indicates the 
 limit obtained without the constraints from the searches for neutralinos. 
 The result obtained with the data collected at $\sqrt{s}$ up to 189\,GeV
 is shown for comparison. 
 Higher $m_0$ values give similar limits.}
 \label{mchi_limit}
 \end{figure}

From the negative outcome of chargino and neutralino searches, 
a lower limit on the mass of the lightest 
neutralino can be derived as a function of \tanb, as shown in 
Fig.~\ref{mchi_limit} for $m_0=500$\,\Gcsq.
Neutralino searches contribute mostly at low $\tanb$ and 
play an essential r\^ole in determining  
the lowest value of 37.2\,\Gcsq, 
obtained for $\tanb=1$, $M_2\simeq 62$\,\Gcsq\ and $\mu\simeq -72$\,\Gcsq, 
which remains valid also for larger $m_0$ values. 
For smaller $m_0$ values, the loss of sensitivity of chargino and neutralino 
searches is recovered by slepton searches, as discussed in Ref.~\cite{chaneu97}. 
 A scan performed over the relevant parameter space, including the deep gaugino 
 region for large $\tan\beta$ and for mass degenerate charginos and sneutrinos, 
 shows that the lower limit on the LSP mass of 37.2\,\Gcsq, obtained for \tanb=1 
 at large $m_0$, holds for all values of $m_0$.

As discussed in Ref.~\cite{efgos}, radiative corrections to chargino and 
neutralino masses modify the relation between such masses and therefore 
the interplay between chargino and neutralino searches. 
In the region where the limit is found, 
these corrections lower the LSP mass limit value by about 1~$\Gcsq$. 
A further source of theoretical uncertainty is represented by 
the GUT relation between $M_1$ and $M_2$.  
Higher--order corrections to the one--loop formula used here have been 
estimated with ISAJET~\cite{isajet} to be of the order of $\pm$3\%, and affect 
by the same amount the lower limit on the LSP mass.

\subsection{Constraints from Higgs boson searches}

\begin{figure}[p]
\begin{center}
\begin{tabular}{c}
\hspace{0.cm}\mbox{\epsfxsize=1.00\hsize\epsfbox{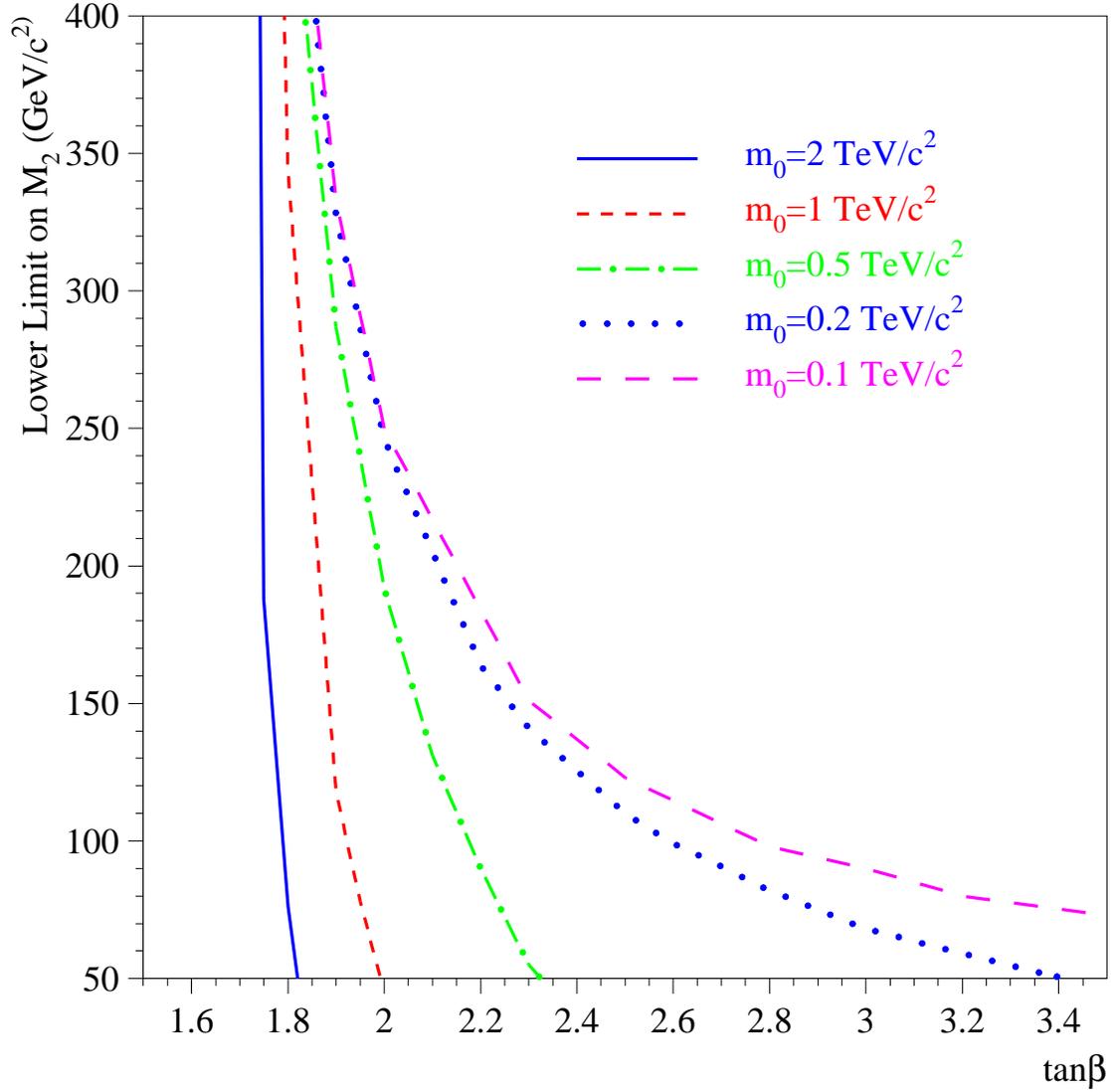}}
\end{tabular}
\caption{\small  
Lower limit on $M_2$ as a function of \tanb, 
as determined from the result of Higgs boson searches, 
for five values of $m_0$.}
\label{M2_vs_tb}
\end{center}
\end{figure}

To derive the results discussed in this section, the analysis presented in 
Ref.~\cite{chaneu97} was updated with the inclusion of the results 
of the ALEPH searches for Higgs bosons~\cite{higgs202}. 
For a given set of \tanb, $m_0$ and $M_2$ values, the largest 
predicted Higgs boson mass is determined with a large value for $\mA$, 
2\,\Tcsq, and maximal effect from stop mixing by varying the combination
$\tilde{\At} = \At-\mu\cot\beta$. For large \mA, the lighter scalar neutral 
Higgs boson becomes standard-model like. It is therefore  
sufficient to compare the largest predicted $\mh$ to the 
standard model Higgs boson mass lower limit of 107.7\,\Gcsq, as determined 
in Ref.~\cite{higgs202}.
The loophole which might have arisen from possible Higgs boson decays into 
stop pairs~\cite{chaneu97} is avoided by the recent parameter--independent 
stop mass lower limit of 63~$\Gcsq$~\cite{sguazzoni}.
The lower limit on $M_2$ obtained in this way is displayed in 
Fig.~\ref{M2_vs_tb} as a function of \tanb, for various values of $m_0$. 
As expected, the limit is strongest for low values of \tanb\ and $m_0$. 
For instance, $M_2$ is larger than 110\,\Gcsq\ for $\tanb=2.5$ and
$m_0 = 200\,\Gcsq$. 

Although weaker as $m_0$ increases, the constraints on $M_2$ are still 
sufficient to improve on the LSP mass lower limit discussed in Section~3.4. 
For instance, for $m_0=1\,\Tcsq$ the limit is 47.5\,\Gcsq\ 
(obtained for $\tanb\simeq 2$); it is 46\,\Gcsq\ for $m_0=2\,\Tcsq$ 
(obtained for $\tanb\simeq 1.8$).

\begin{figure}[ht]
\begin{center}
\begin{tabular}{c}
\hspace{0.cm}\mbox{\epsfxsize=1.00\hsize\epsfbox{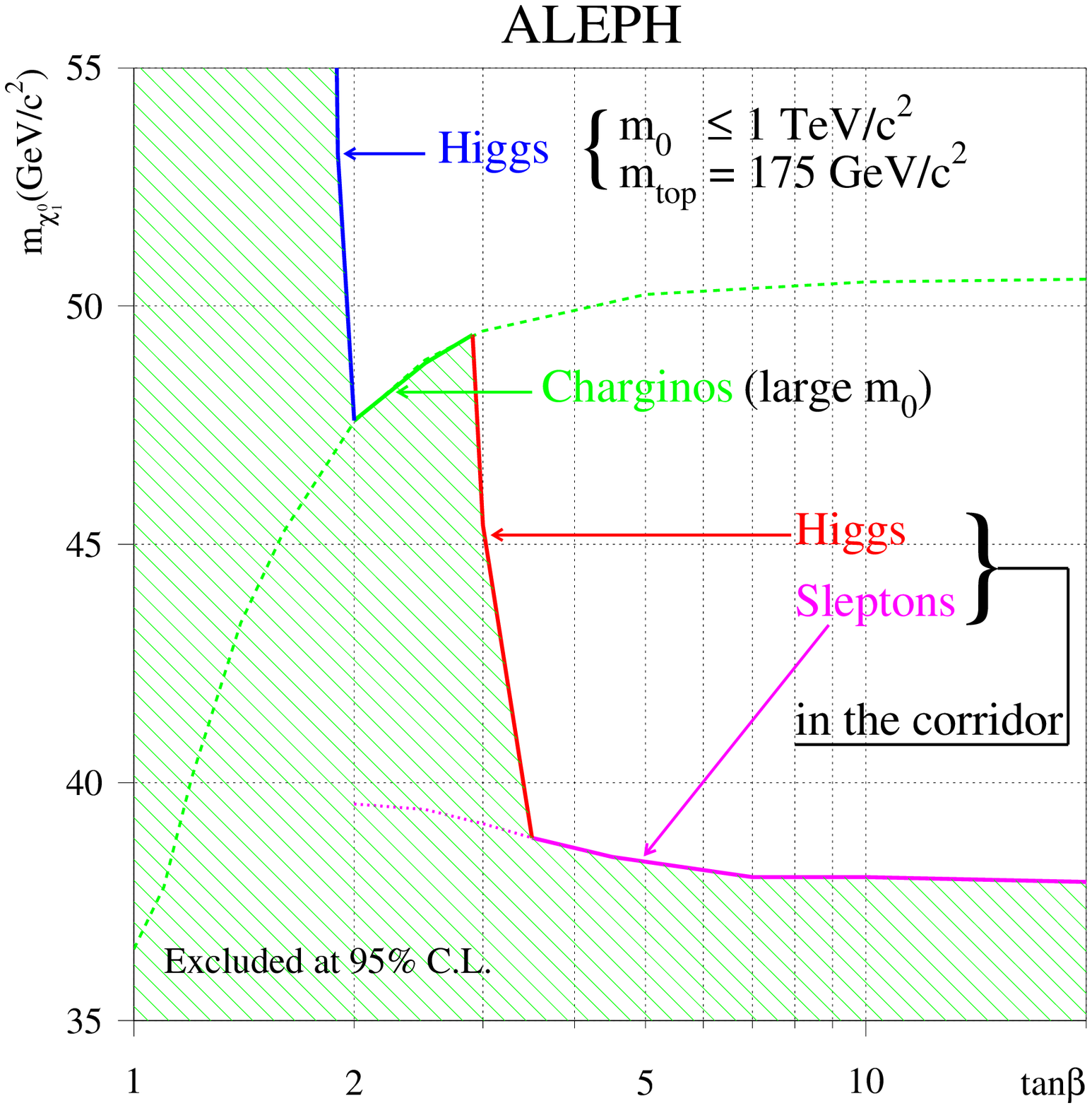}}
\end{tabular}
\caption{\small Lower limit on \Mchi\, as a function of \tanb, from, right to 
left, slepton searches in the corridor ($m_{\Pcha}\simeq m_{\PSnu}$), 
Higgs boson searches in the corridor, 
chargino searches for large sfermion masses, and Higgs boson searches. 
The limits from Higgs boson searches are valid for $\mt\!=\!175\,\Gcsq$
and for $m_0$ not exceeding 1\,\Tcsq. The dashed curve indicates the limit 
from chargino and neutralino searches for large $m_0$.  The dotted curve 
shows the limit in the corridor reached without the Higgs boson searches.}
\label{chi_lim}
\end{center}
\end{figure}

In the small $m_0$ regime, the LSP mass lower limit discussed in Section~3.4 
depends on the extent to which the slepton searches cover the so--called {\it corridor}, 
a subset of model parameters giving sneutrinos almost degenerate in mass 
with the lightest charginos. 
The constraints from Higgs boson searches provide additional coverage 
of the corridor at small $\tanb$, as depicted in Fig.~\ref{chi_lim}.
While for $\tanb\!>3.5$ the Higgs boson constraints do not bring any 
improvement, the lower limit on $\Mchi$ 
set in the corridor by Higgs boson searches is larger than that set 
by slepton searches for $2.85\!<\!\tanb\!<\!3.5$.

For $\tanb < 2.85$, the limit on \Mchi\ set in the corridor 
by Higgs boson searches is no longer 
the absolute limit, since it exceeds that obtained from chargino searches 
for large $m_0$ values. This latter limit might degrade  
as $m_0$ is reduced, but the Higgs boson searches confine the charginos more 
and more in the higgsino region (Fig.~\ref{M2_vs_tb}), thus rendering the 
deleterious influence of light sleptons and light sneutrinos less and less 
important. As a result, the limit from chargino searches at large $m_0$ and 
$\tanb < 2.85$ remains robust for any smaller $m_0$ value, and is even superseded
by the limit arising from Higgs boson searches for $\tanb\!<\!1.95$. 

The results obtained with the constraints from Higgs boson searches are
however quite sensitive to the top quark mass. A value of 175\,\Gcsq\ was used 
to derive the results reported in this section, {\it i.e.}, an LSP mass lower limit 
of 38\,\Gcsq, irrespective of \tanb, and of 45\,\Gcsq\ for $\tanb < 3$
(Fig.~\ref{chi_lim}). If a value of 180\,\Gcsq\ is chosen instead, 
the absolute lower limit of 38\,\Gcsq\ still holds,
but the limit for $\tanb < 3$ is reduced to 40\,\Gcsq. If, in addition, 
a value as large as 2\,\Tcsq\ (instead of 1\,\Tcsq) is allowed for $m_0$,  
the Higgs boson searches no longer improve at low \tanb\ on the limit of 
37.2\,\Gcsq\ deduced from the chargino and neutralino searches for large 
sfermion masses.

\subsection{Interpretation in Minimal Supergravity}
{\label{sec:sugra}}

\begin{figure}[hp]
\begin{center}
\mbox{\epsfig{figure=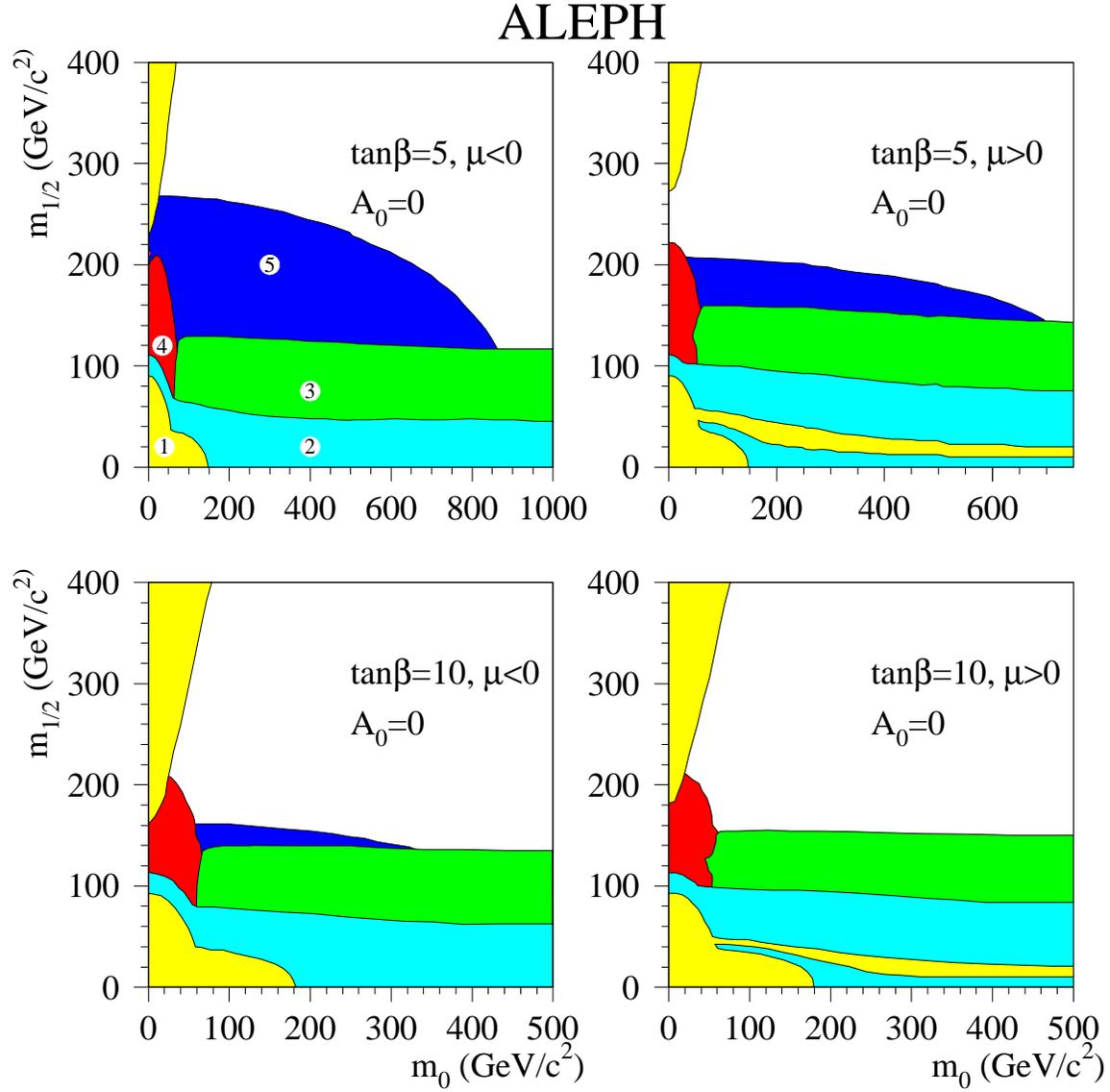,height=16cm}}
\end{center}
\caption{\small Minimal supergravity scenario: domains of the 
$(m_0,m_{1/2})$ plane excluded for 
$\tan\beta=5$ and 10 and for $A_0=0$.
Region 1 is theoretically forbidden. The other regions are excluded by
the Z width measurement at LEP1 (2), chargino (3), slepton (4) searches 
and Higgs boson searches at $\sqrt{s}\leq 201.6$\,GeV (5).
}
\label{fig_SUGRA}
\end{figure}

The interplay among the searches for sleptons, charginos and  Higgs bosons,
and the $\mathrm{Z}$ width measurement at LEP1~\cite{EW}, is shown 
in Fig.~\ref{fig_SUGRA} as exclusion domains in the $(m_0,m_{1/2})$ plane  
for $\tanb=5$ and 10, for $\mu<0$ and $\mu>0$, and for $A_0=0$;  
the top quark mass was set to 175 $\Gcsq$. 

The scan of the $(m_0,m_{1/2})$ plane allows constraints to be derived 
on the mass of the lightest neutralino. 
As already noticed in Ref.~\cite{LEPsugra}, it is possible 
to find configurations where the analyses 
discussed in this letter lose their sensitivity; these loopholes
open up when the lightest \PStau\ is almost degenerate with the $\PChiz{1}$.
To obtain the results reported in this section, 
these configurations were avoided by the requirement that 
$m_{\PStau}\!-\!\Mchi$ be larger than $5\,\Gcsq$.

\begin{figure}[hp]
\begin{center}
\mbox{\epsfig{figure=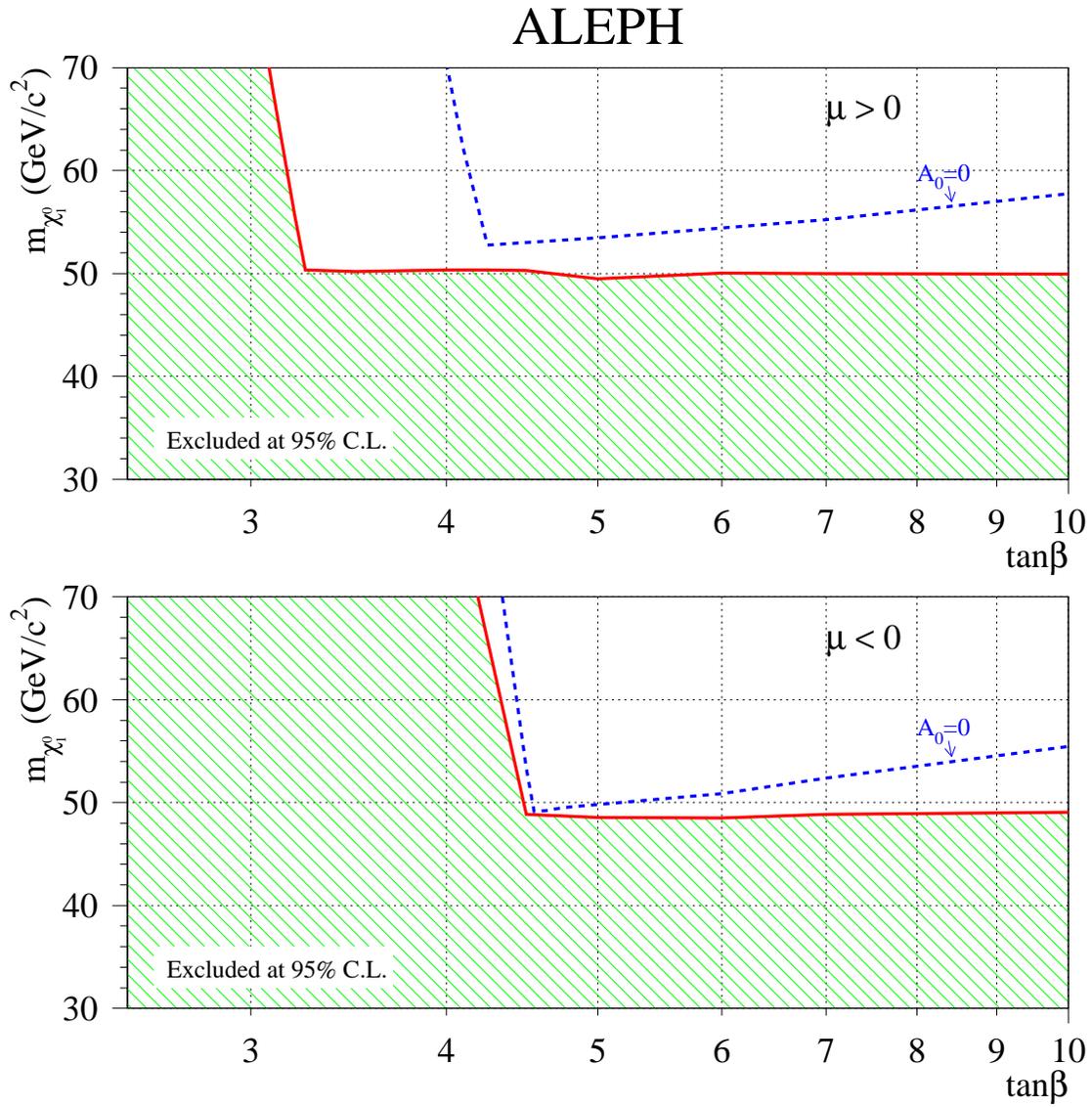,height=16cm}}
\end{center}
\caption{\small Minimal supergravity scenario: lower limit on the 
LSP mass as a function of \tanb. The dashed line shows the 
result obtained with $A_0=0$.}
\label{mchi_sugra}
\end{figure}

The lower limit on \Mchi\ as a function of \tanb\ is shown in 
Fig.~\ref{mchi_sugra} for $A_0=0$ and for both $\mu<0$ and $\mu>0$. 
There is little structure in the limit for both signs of $\mu$. Small \tanb\ 
values are excluded by the negative result of Higgs boson searches, 
whereas for larger \tanb\ values the limit is determined mostly by 
chargino searches.
The lowest allowed value for \Mchi\ is always found at large $m_0$;
it has been calculated for $m_0=1\,\Tcsq$. 
Altogether, a $\PChiz{1}$ mass lower limit of 49\,\Gcsq\ is set for $A_0=0$. 
It is reached for $\tanb \sim 4.5$ and $\mu<0$.

The impact of a non-vanishing $\Azero$ value was studied by scanning the  
range allowed by theoretical constraints and by stop searches.
The lower limit on \Mchi\ as a function of \tanb\ obtained from the $\Azero$
 scan is also shown in Fig.~\ref{mchi_sugra}.
Altogether, the $\PChiz{1}$ mass lower limit decreases to 48\,\Gcsq, a value 
reached for $\tanb \sim 4.5$ and $\mu<0$. 
The lower limit on $\Mchi$ is therefore substantially improved when 
the additional constraints of minimal supergravity are considered.

\section{Conclusions}

\label{sec:conclusions}
The previously published ALEPH searches for 
sleptons, squarks, charginos and neutralinos
have been updated with the data collected during 1998 and 1999 at 
centre-of-mass energies up to 201.6\,GeV, corresponding to an overall 
integrated luminosity of about 410~$\pbinv$. 
In all topologies, the numbers of candidate events  
observed are consistent with the background expected from standard   
processes. When interpreted in the framework of the MSSM, this negative 
outcome allows improved 95\% C.L. lower limits on slepton and squark masses 
to be set as summarized in Table~4.

\renewcommand{\arraystretch}{1.2}
\begin{table}[ht]
\caption{ \small Lower limits at 95\% C.L. on squark and slepton masses. 
For selectrons and sleptons, $\tan\beta=2$, $\mu=-200$~\Gcsq. 
For sleptons, a common mass at the GUT scale is assumed. 
All masses and mass differences are in $\Gcsq$.
}
\label{limits}
 \begin{center}
 \begin{tabular}{|l|c|c|}
\hline
 Particle         & Limit & Conditions of validity \\
\hline
\hline
 selectron & 92 & $\dm\!>\!10$~ \\
\hline
 smuon     & 85 & $\dm\!>\!10$~, 
                   \mbox{$\tilde{\mu}\!\to\!\mu\PChiz{1}$} \\
\hline
 stau      & \raisebox{-3mm}[0mm][-3mm]{68} & $\dm\!>\!10$~,  
                    \mbox{$\tilde{\tau}\!\to\!\tau\PChiz{1}$}, \\
           &     & worst case mixing \\ 
\cline{2-3}
                  & 70 & $\dm\!>\!10$~, 
                   \mbox{$\tilde{\tau}\!\to\!\tau\PChiz{1}$}, 
                   \PStaur \\
\hline
 slepton   & 93 & $\dm\!>\!10$~ \\ 
\cline{2-3}
                  & 70 & any $\dm$  \\
\hline\hline
 stop      & 83 & $\PSt\!\to\!\mathrm{c}\PChiz{1}$,
                               $6 < \dm < 40$ \\ 
\cline{2-3}
                  & 88 & $\mathrm\PSt\!\rightarrow\!\mathrm{b}
         \ell\widetilde{\nu}$,
                       $\dm > 10$; \\
\hline
 sbottom   & 91 & $\PSb\!\to\!\mathrm{b}\PChiz{1}$, $\dm > 8$,
                       \PSbl \\
\hline
 degenerate  &   \raisebox{-3mm}[0mm][-3mm]{97} 
           & $\tilde{\mathrm{q}}\!\to\!\mathrm{q}\PChiz{1}$, $\dm > 6$,  \\
 squarks &        &        $\tan\beta=4$, $\mu=-400$ \\ 
\hline
\end{tabular}
\end{center}

\end{table}
\renewcommand{\arraystretch}{1.}

Chargino pair production and neutralino associated production are excluded up 
to the kinematic limit in a significant fraction of the MSSM parameter space. 
For $m_0>500$\,\Gcsq, a lower limit on \Mchi\ of 37\,\Gcsq\ is obtained, 
independent of \tanb. It has been
verified that this limit remains valid for all $m_0$ values if 
the constraints from slepton searches are taken into account, for 
a negligible mixing in the stau sector. 

The negative results of the searches for Higgs bosons reported in 
Ref.~\cite{higgs202} further constrain the MSSM parameter space.
In particular, the lower limit on the $\PChiz{1}$ mass is increased to
45\,\Gcsq\ for $\tanb\!<\!3$ and to 38\,\Gcsq\ for any \tanb,
assuming $\mt\!=\!175$\,\Gcsq\ and $m_0\!<\!1$\,\Tcsq. 

The results have also been interpreted in 
the framework of minimal supergravity,
including one-loop radiative corrections in the calculation of
chargino and neutralino masses.
The domains excluded in the $(m_0,m_{1/2})$ plane are significantly 
larger than those reported by the D0 collaboration from the negative results 
of searches for gluinos and squarks at the Tevatron~\cite{D0mles}. 
The resulting lower limit on \Mchi\ is~49\,\Gcsq\ for $\Azero\!=\!0$ and 
$m_0\!<1\!\,\Tcsq$. It is reduced to 48\,\Gcsq\ when $\Azero$ is 
varied, provided that the parameter configurations leading to 
$m_{\PStau}\!-\!\Mchi < 5\,\Gcsq$ are ignored.

These results improve significantly on those obtained at lower energies by 
ALEPH~\cite{chaneu97} and by the L3~\cite{L3189}, OPAL~\cite{OPAL189} 
and DELPHI~\cite{DELPHI183} collaborations. 
Less general searches for associated production of charginos and neutralinos 
have also been reported by the CDF~\cite{CDF} and D0~\cite{D0} collaborations.

\subsection*{Acknowledgements}
It is a pleasure to congratulate our colleagues from the accelerator divisions
for the successful operation of LEP2. 
We are indebted to the engineers and technicians in all our institutions
for their contribution to the excellent performance of ALEPH.
Those of us from non-member states wish to thank CERN for its hospitality
and support.

\mbox{~}
\vfill

\end{document}